\ifx\documentclass\undefined
\documentstyle[12pt]{article}  %% For those without LaTeX2e

\def\rz{\ifmmode {I\hskip -3pt R} \else {\hbox {$I\hskip -3pt R$}}\fi}
\def\cz{\ifmmode {C\hskip -4.8pt\vrule height5.8pt\hskip 6.3pt} \else
       {\hbox {$C\hskip -4.8pt\vrule height5.8pt\hskip 6.3pt$}}\fi}

\newcommand{\gA}{{\bf A}} % Vector potential
\newcommand{\gB}{{\bf B}} % Magnetic induction

\newcommand{\gj}{{\bf j}} % Current
\newcommand{\gn}{{\bf n}} % the unit vector n
\newcommand{\nul}{{\bf 0}}% The null vector
\newcommand{\gp}{{\bf p}} % momentum variable 
\newcommand{\gq}{{\bf q}} % momentum variable 
\newcommand{\gr}{{\bf r}} % position of an electron
\newcommand{\gR}{{\bf R}} % position of a nucleus
\newcommand{\cA}{{\cal A}}% generic magnetic field
\newcommand{\cE}{{\cal E}}

\newcommand{\BS}[1]{{\cal E}_{#1}}      % Brown-Ravenhall form
\newcommand{\cH}{{ \cal H}_+} % 1-particle Hilbert space
\newcommand{\cHm}{{\cal H}_-} 
\newcommand{\bHN}[1]{{{\cal H}_{N,#1}}} % N-particle Hilbert space
\def\tr{\mathop{{\rm tr}} \nolimits} % trace
\else
\documentclass[12pt]{article}  %% For those with LaTeX2e

\IfFileExists{amssymb.sty}
{ %%For those with LaTeX2e and amssymb.sty
  \usepackage{amssymb}
  \newcommand{\rz}{\mathbb{R}} % Reelle Zahlen
  \newcommand{\cz}{\mathbb{C}} % Komplexe Zahlen
}
{ %%For those with LaTeX2e but without amssymb.sty
  \def\rz{\ifmmode {I\hskip -3pt R} \else {\hbox {$I\hskip -3pt R$}}\fi}
  \def\cz{\ifmmode {C\hskip -4.8pt\vrule height5.8pt\hskip 6..3pt} \else
  {\hbox {$C\hskip -4.8pt\vrule height5.8pt\hskip 6.3pt$}}\fi}
}

\newcommand{\gA}{\mathbf{A}} % Vector potential
\newcommand{\gB}{\mathbf{B}} % Magnetic induction

\newcommand{\gj}{\mathbf{j}} % Current
\newcommand{\gn}{\mathbf{n}} % the unit vector n
\newcommand{\nul}{\mathbf{0}}% The null vector
\newcommand{\gp}{\mathbf{p}} % momentum variable 
\newcommand{\gq}{\mathbf{q}} % momentum variable 
\newcommand{\gr}{\mathbf{r}} % position of an electron
\newcommand{\gR}{\mathbf{R}} % position of a nucleus
\newcommand{\cA}{\mathcal{A}}% generic magnetic field
\newcommand{\cE}{\mathcal{E}}

\newcommand{\BS}[1]{\mathcal{E}_{#1}}      % Brown-Ravenhall form
\newcommand{\cH}{\mathcal{H}_+} % 1-particle Hilbert space
\newcommand{\cHm}{\mathcal{H}_-} 
\newcommand{\bHN}[1]{{\mathcal{H}_{N,#1}}} % N-particle Hilbert space
\def\tr{\mathop{\mathrm{tr}} \nolimits} % trace
\fi

\def\qed{\hbox {\hskip 1pt \vrule width 6pt height 6pt depth 1.5pt
        \hskip 1pt}}% Beweisende

\def\atp{\bigwedge\limits} % antisymmetrisches Tensorprodukt
\newcommand{\balpha}{\mbox{\boldmath$\alpha$}}% vector of alpha matrix
\newcommand{\bsigma}{\mbox{\boldmath$\sigma$}}% vector of sigma matrix
\newcommand{\bomega}{\mbox{\boldmath$\omega$}}% omega vector

\newcommand{\iint}{{\int\!\!\int}} % doubleintegral

\newtheorem{theorem}{Theorem}
\newtheorem{definition}{Definition}
\newenvironment{proof}{{\it Proof. \/}}{$\qed$}

\title{Stability and Instability of Relativistic Electrons in
  Classical Electromagnetic Fields\footnotetext{\copyright 1996 by the
    authors.  Reproduction of this article, in its entirety, by any
    means is permitted for non-commercial purposes}}

\author{Elliott H.~Lieb\\
        Departments of Mathematics and Physics\\
        Princeton University\\
        Princeton, NJ 08544-0708
\and
        Heinz Siedentop\\
        Matematisk institutt\\
        Universitetet i Oslo\\
        Postboks 1053\\
        N-0316 Oslo
\and 
        Jan Philip Solovej\\
        Institut for matematiske fag\\
        Aarhus universitet\\
        Ny Munkegade\\
        DK-8000 \AA rhus C
}
\date{Dedicated to Bernard Jancovici on his 65th birthday.} 
\begin{document}
\maketitle
\begin{abstract}
  The stability of matter composed of electrons and static nuclei is
  investigated for a relativistic dynamics for the electrons given by
  a suitably projected Dirac operator and with Coulomb interactions.
  In addition there is an arbitrary classical magnetic field of finite
  energy. Despite the previously known facts that ordinary
  nonrelativistic matter with magnetic fields, or relativistic matter
  without magnetic fields is already unstable when $\alpha$, the fine
  structure constant, is too large it is noteworthy that the
  combination of the two is still stable {\it provided} the projection
  onto the positive energy states of the Dirac operator, which {\it
    defines} the electron, is chosen properly.  A good choice is to
  include the magnetic field in the definition. A bad choice, which
  always leads to instability, is the usual one in which the positive
  energy states are defined by the free Dirac operator. Both
  assertions are proved here.
\end{abstract}

\section{Introduction}\label{s1} 
The stability of matter concerns the many-electron and many-nucleus
quantum mechanical problem and the question whether the ground state
energy is finite (stability of the first kind). If so, is it bounded
below by a constant (which is independent of the position of the
nuclei) times the number of particles (stability of the second kind)?
The linear lower bound is important for thermodynamics, which will not
exist in the usual way without it.

The first positive resolution of this problem for the nonrelativistic
Schr\"o\-din\-ger equation was given by Dyson and Lenard
\cite{DysonLenard1967I,DysonLenard1967II} and approached differently
by Federbush \cite{Federbush1975}.  The constant, i.e., the energy per
particle, was considerably improved by Lieb and Thirring in
\cite{LiebThirring1975,LiebThirring1976}.  Following that, the
stability of a relativistic version of the Schr\"odinger equation (in
which $\gp^2$ is replaced by $\sqrt{\gp^2 +m^2}$) was proved by Conlon
\cite{Conlon1984} and later improved by Lieb and Yau
\cite{LiebYau1988} who showed that matter is stable in this model if
and only if the fine structure constant $\alpha$ is small enough and
if $Z\alpha \leq 2/\pi$. (See \cite{LiebYau1988} for a historical
account up to 1995.) A recent result of Lieb, Loss, and Siedentop that
we shall use is in \cite{Liebetal1996} and is discussed in Section
\ref{s2}.

In these works the nuclei are fixed in space because they are very
massive and because we know that the nuclear motion is largely
irrelevant for understanding matter. In other words, if nuclear motion
were the only thing that prevented the instability of matter then the
world would look very different from what it does. We continue this
practice here.
 
There is, however, a more important quantity that requires some
attention, namely magnetic fields. It was noted that the action of
such fields on the translational degrees of freedom of the electrons
$\gp\rightarrow\gp+e\gA$, can lower the energy only by an
inconsequential amount. This is a kind of diamagnetic inequality. On
the other hand, spin-magnetic field interaction (in which
$(\gp+e\gA)^2$ is replaced by the Pauli operator
$[\mbox{\boldmath$\sigma$}\cdot(\gp+e\gA)]^2
=(\gp+e\gA)^2+e\mbox{\boldmath$\sigma$}\cdot\gB$ can cause
instability. The energy is then unbounded below if arbitrarily large
fields are allowed, but this is so only because the energy of the
magnetic field has not been taken into account. Does the field energy,
$(8\pi)^{-1} \int B^2$, insure stability?  This question was raised
for the nonrelativistic case in \cite{Frohlichetal1986} and finally
settled in a satisfactory manner in \cite{Liebetal1995} (see also
Bugliaro et al. \cite{Bugliaroetal1996} and Fefferman
\cite{Fefferman1996}). The upshot of this investigation is that
stability (of both first and second kinds) requires a bound on both
$\alpha$ and on $Z \alpha^2$.

Other related results are the stability of non-relativistic matter
with a second quantized, ultra-violet cut-off photon field
(Fr\"ohlich et al. \cite{Feffermanetal1996}).

Both the passage to relativistic kinematics (which, in quantum
mechanics, means that both the kinetic energy and the Coulomb
potential scale with length in the same way, namely like an inverse
length) and the introduction of the nonrelativistic Pauli operator
require a bound on $\alpha$ and on $Z$ for stability. The combination
of the two might be expected to lead to disaster. We find, however,
that it does not necessarily do so! 

{\it Our main result is that matter is indeed stable with a suitably
  defined relativistic kinematics.} This is shown in Section \ref{s2}.

The proper way to introduce relativistic kinematics for spin-1/2
particles is via the Dirac operator, but this is unbounded below. A
resolution of this problem, due to Dirac, is to permit the electrons
to live only in the positive energy subspace of the Dirac operator.
This idea was further pursued by Brown and Ravenhall
\cite{BrownRavenhall1951} (see also Bethe and Salpeter in their
Handbuch article \cite{BetheSalpeter1957}) to give a quantitative
description of real atoms.

There are, however, other Dirac operators (which include electromagnetic
potentials) whose positive subspace can be used to define the space in
which the electrons can live.  (To avoid confusion, let us note that
the Hamiltonian is formally always the same and includes whatever
fields happen to be present. The only point to be resolved is what
part of the one-particle Hilbert space is allowed for electrons.) The
review articles of Sucher
\cite{Sucher1980,Sucher1984,Sucher1987}  can be consulted here.  
These choices have also been used in quantum chemistry and other
practical calculations, see, e.g., \cite{IshikawaKoc1994,Jensenetal1996}.

All of these choices have in common that there is no creation of
electron-positron pairs explaining the name ``no-pair Hamiltonian''
for the resulting energy operator. (Note that we could also treat
positrons or a combination of electrons and positrons interacting by
Coulomb forces in a similar way.)

There are three obvious choices to consider. One is the {\it free} Dirac
operator.  {\it This always leads to instability of the first kind
when a magnetic field is added} unless the particle number is held to
some small value (see Section \ref{s3}). Note also that this choice
leads to a non-gauge invariant model: multiplication of a state with
the factor $\exp(i\phi(\gr))$ for a non-constant gauge is not allowed,
since it leads out of the positive spectral subspace.

Remarkably, the Dirac operator that includes the magnetic field always
gives stability, if $Z$ and $\alpha$ are not too large, as in the two
cases (relativistic without magnetic field and nonrelativistic with
magnetic field) mentioned above (see Section \ref{s2}). This model
{\it is} gauge invariant.

The third choice which, indeed, is sometimes used, is to include {\it
both} the one-body attractive electric potential of the nuclei and the
magnetic field in the definition of the Dirac operator that defines
the positive subspace. If this is done then the question of stability
is immediately solved because the remaining terms in the Hamiltonian
are positive, and hence the total energy is ipso facto positive.  This
choice, which is important but trivial in the context of this present
inquiry, will not be mentioned further.

Oddly, the instability proof given in Subsection \ref{ss32} is much
more complicated than the stability proof (Section \ref{s2}). This
reverses the usual situation.

A preliminary  version of this paper appeared in \cite{WT}; the present
version is to be regarded as the original one (as stated in \cite{WT}) and
contains several significant corrections to the preliminary text in
\cite{WT}.  In particular, the proof of Theorem 2 and the first half of the
proof of Theorem 1 have been corrected and simplified.

\section{Basic Definitions \label{s1a}}

The energy of $N$ relativistic electrons in the field of $K$ nuclei
with atomic numbers $Z_1,...,Z_K \in \rz_+$ located at
$\gR_1,...,\gR_K\in \rz^{3}$ which are pairwise different
in a magnetic field $\gB=\nabla\times \gA$ in the state $\Psi$
is---following the ideas of Brown and Ravenhall
\cite{BrownRavenhall1951}---
\begin{equation}\label B
  \BS{\cA}[\Psi,\Psi] := (\Psi,(\sum_{\nu=1}^N D_\nu(\gA) + \alpha
  V_c)\Psi) + \frac1{8\pi}\int_{\rz^3}B(\gr)^2d\gr.
\end{equation}
Here $D_\nu(\gA) :=\mbox{\boldmath$\alpha$} \cdot(- i\nabla_\nu +
e\gA(\gr_\nu)) +m \beta$ is the Dirac operator with vector potential
$\gA$. Furthermore, 
\begin{equation}
\label{e:coulomb}
V_c:=-\sum_{\nu=1}^N\sum_{\kappa=1}^K\frac{Z_\kappa}{|\gr_\nu -
\gR_\kappa|} + \sum_{\mu,\nu=1\atop \mu<\nu}^N
\frac{1}{|\gr_\mu-\gr_\nu|} + \sum_{\kappa,\lambda=1 \atop
\kappa<\lambda}^K \frac{Z_\kappa Z_\lambda}{|\gR_\kappa-\gR_\lambda|} 
\end{equation}
is the Coulomb interaction between the particles, and $B(\gr):=
  |\nabla\times\gA(\gr)|$ is the modulus of the magnetic field.
  Planck's constant divided by $2\pi$ and the velocity of light, are
  taken to be one in suitable units.  The fine structure constant
  $\alpha$ equals $ e^2$, where $-e$ is the electron
  charge. Experimentally, $\alpha$ is about 1/137.037. The mass of the
  electron is denoted by $m$. The $4\times4$ matrices
  {\boldmath$\alpha$} and $\beta$ are the four Dirac matrices in
  standard representation, namely
$$ \mbox{{\boldmath$\alpha$}}=
  \left(\begin{array}{cc} 0&\mbox{{\boldmath$\sigma$}}\\
\mbox{{\boldmath$\sigma$}}&0\end{array}\right),\ 
\sigma_1=\left(\begin{array}{rr} 0&1 \\ 1&0 \end{array}\right),\ 
\sigma_2=\left(\begin{array}{rr} 0&-i \\ i&0 \end{array}\right),\
\sigma_3=\left(\begin{array}{rr} 1&0 \\ 0&1 \end{array}\right),
$$
and
$$
\beta = \left(\begin{array}{rrrr}
1&0&0&0\\ 0&1&0&0\\ 0&0&-1&0\\ 0&0&0&-1\end{array}\right).$$

Finally, the state $\Psi$ should have finite kinetic energy, i.e., it
should be in the Sobolev space
$H^{1/2}[(\rz^3\times\{1,2,3,4\})^{N}]$, and should also be in the
electronic Hilbert space of antisymmetric spinors
\begin{equation}\label{HS}
\bHN{\cA}
:=\atp_{\nu=1}^N \cH \ ,
\end{equation}
where $\cH$ is the {\it positive spectral subspace of the
Dirac operator} $D(\cA)$ and where $\cA$ is some vector potential to
be chosen later. The vector potential $\cA$ serves to {\it define} the
positive subspace. Two choices will be considered here. One is $\cA
=\nul$, in which case we are talking about the free Dirac
operator. This choice, or model, goes back to Brown and Ravenhall
\cite{BrownRavenhall1951}. 
As we shall see in Section \ref{s3},
the resulting energy functional---apart from being not gauge
invariant---is not bounded from below.   A
natural modification of the model, namely to take $\cA:=\gA$ is not
only gauge invariant, but will also turn out to be 
stable of the second kind (see Section \ref{s2}). 

The quantity of interest is the lowest possible energy 
$$E_{N,K}:=\inf \BS{\cA}$$
where the infimum is taken over all allowed states $\Psi$, all allowed
vector potentials $\gA$, and over all pairwise different nuclear
positions $\gR_1,...,\gR_K$.

In the case of a single nucleus without a magnetic field, the energy
form $\BS{\nul }$ was shown in \cite{Evansetal1996} to be bounded from
below, if and only if $\alpha Z\leq \alpha
Z_C:=2/(\pi/2+2/\pi)>2/\pi$, which corresponds to $Z\approx 124$. We
will not be able to reach this value in the general case of many
nuclei and when the electron state space is not determined by the free
Dirac operator.  The reason is that special techniques were used in
\cite{Evansetal1996} to handle the one-nucleus case; 
these techniques took advantage of the weakening of the Coulomb
singularities caused by the fact that states in $\cH$ cannot
be localized in space arbitrarily sharply.  Unfortunately, we do not
know how to implement this observation with magnetic fields and many
nuclei.

\section{Stability with the Modified Projector \label{s2}}
Our proof of the stability of matter when the vector potential $\gA$
is included in the definition of the positive energy electron states
will depend essentially on three inequalities:
\begin{description}
\item[BKS inequality:] For any self-adjoint operator $X$, the negative
  (positive) part, $X_\mp$ is defined to be $ (|X|\mp X)/2$. Given two
  non-negative self-adjoint operators $C$ and $D$ such that
  $(C^2-D^2)_-^{1/2}$ is trace class, we have the trace inequality
\begin{equation}\label{e2} 
\tr(C-D)_- \leq \tr(C^2-D^2)_-^{1/2}.
\end{equation} 
This is a special case of a more general inequality of Birman,
Koplienko, and Solomyak \cite{Birmanetal1975}; in particular, the
number 2 in (\ref{e2}) can be replaced by any $p>1$. A proof for the
special case of the inequality needed here is given in Appendix
\ref{a:1}.
\item[Stability of relativistic matter:] 
On $\atp_{\nu=1}^N(H^{1/2}(\rz^3)\otimes \cz^q)$, the fermionic
  Hilbert space, we have
\begin{equation}\label{e3}
    \sum_{\nu=1}^N |-i\nabla_\nu-\cA| + \tilde\alpha V_c\geq0,
\end{equation} 
(where $|\cdots|$ means $\sqrt{ (\cdots)^2}$\ ) for all vector fields
$\cA:\rz^3\rightarrow\rz^3$ with, e.g., square integrable gradients,
if
\begin{equation}\label{e:3a}
    1/\tilde\alpha \geq 1/\tilde\alpha_c := (\pi/2)Z + 2.2159
    q^{1/3}Z^{2/3} + 1.0307 q^{1/3}
\end{equation} 
and $Z_1,...,Z_K\leq Z$. 

We wish to use this inequality for 4-component spinors, i.e., $q=4$.
However, we are interested in the subspace $\bHN{\cA}$ in which the
particles are restricted to the positive energy subspace of the Dirac
operator, $D(\cA)$.  Although $q=4$, the `effective' $q$ is really 2,
and the analysis in Appendix \ref{a:2} is our justification for
this. The only thing that really counts in deriving (\ref{e3}) is the
bound on the reduced one-body density matrix $\gamma$ mentioned in
Appendix \ref{a:2}.

The stability of the relativistic Hamiltonian
(\ref{e3}) was first shown by Conlon \cite{Conlon1984} for
$\cA=\nul$. The best currently available constants with $\cA=\nul$ are
in \cite{LiebYau1988} while (\ref{e:3a}), which is taken from
\cite{Liebetal1996}, is the best known result for general $\cA$.
\item[Semi-classical bound:] Given a positive constant $\mu$, a
real vector field $\cA$ with, e..g., square integrable gradients, and a
real-valued function $\varphi\in L^2(\rz^3)$ the inequality
\begin{equation}
\label{e4} \tr[(-i\mu \nabla-\cA)^2 - \varphi]_-^{1/2}
\leq \frac{ L_{1/2,3} } {\mu^3}\int_{\rz^3} \varphi_+^2 
\end{equation}
holds, which is a special case of the Lieb-Thirring inequality (see
\cite{LiebThirring1976,Lieb1984O}). It is known that $L_{1/2,3}\leq
0.06003$. The left side of (\ref{e4}) is simply
$\sum_{j}|\lambda_j|^{1/2}$, where the $\lambda_j$ are the negative
eigenvalues of the operator $[\cdots]$.
\end{description}

As an illustration of the usefulness of the trace estimate (\ref{e2}),
let us combine it with the Lieb-Thirring inequality (\ref{e4}) (or any
other Lieb-Thirring inequality for that matter) to derive some
previously known inequalities. The constants obtainable in this way are
comparable with the best ones known so far:
\begin{description}
\item[Daubechies inequality:]
We begin with a ``relativistic'' inequality that was first proven by
Daubechies \cite{Daubechies1983}. By replacing $\varphi$ by
$\varphi^2$ in (\ref{e4}), we get using (\ref{e2})
\begin{equation}
\tr\left(|-i\nabla -\cA|- \varphi\right)_- 
\leq L_{1/2,3}\int_{\rz^3}\varphi_+^4.
\label{e:4a}
\end{equation}
The constant 0.06003 obtained here should be compared with the number
0.0258 in \cite{Daubechies1983}.
\item[Non-relativistic magnetic stability:] 
A non-relativistic analogue of our main problem is to bound the
form
\begin{equation}
\label{e:nr}
\tilde\cE := (\Psi,(\sum_{\nu=1}^N P_\nu(\gA) + \alpha V_c)\Psi) +
\frac1{8\pi}\int_{\rz^3}B(\gr)^2d\gr
\end{equation}
which was treated in \cite{Liebetal1995}.  Here $P_\nu(\gA)
:=[\mbox{\boldmath$\sigma$} \cdot(- i\nabla_\nu + e\gA(\gr_\nu))]^2$
is the Pauli operator with vector potential $\gA$.

First, we note that $x^2 \geq +\lambda |x| - \lambda^2/4 $ holds.  A
constant in the energy form, however, is irrelevant for checking on
stability of the second kind.  Using (\ref{e3}) it is then enough to
show the positivity of
\begin{equation}
\label{et}
-\tr ( \lambda P(\gA)^{1/2}-\kappa|-i\nabla+e\gA|)_- +
\frac1{8\pi}\int_{\rz^3}B(\gr)^2d\gr
\end{equation}
where we have set $\kappa:=\alpha/\tilde\alpha_c$. 
The trace in this and the next expression are over $L^2(\rz^3)\otimes
\cz^2$.
Using the BKS inequality gives the lower bound
$$
 -\tr [(\lambda^2-\kappa^2)|-i\nabla+e\gA|^2) -e \lambda^2 B(\gr)]_-^{1/2} +
  \frac1{8\pi}\int_{\rz^3}B(\gr)^2d\gr.
$$
Applying the Lieb-Thirring inequality (\ref{e4}) yields 
the following sufficient condition for stability (recall that
$\alpha=e^2$ and that there are {\it two} spin states)
$$
2L_{\frac12,3}\frac{\lambda^4\alpha}{(\lambda^2-\kappa^2)^{3/2}} \leq
\frac{1}{8\pi}.
$$
Optimizing in $\lambda$ gives
$$
2L_{\frac12,3}\frac{16\alpha^2}{3^{3/2}\tilde\alpha_c} \leq \frac{1}{8\pi},
$$ which gives for the physical values $\alpha\approx 1/137.037$ and
$q=2$ a range of stability up to $Z\leq 1096$, which is to be compared
with  $Z\leq 1050$ in \cite{Liebetal1995}..
\end{description}

We turn now to our main result. 

\begin{theorem}\label{t1}
  Let $Z_1,...,Z_K \leq Z < 2/(\pi\alpha)$ and let $\alpha\leq
  \alpha_c$ where $\alpha_c$ is the unique solution of the equation
$$
(16\pi L_{\frac12,3}\, \alpha_c)^{2/3} = 1-\alpha_c^2/\tilde\alpha_c^2,
$$ 
with
$\tilde\alpha_c:=[(\pi/2)Z + 2.2159\cdot2^{1/3}Z^{2/3} + 1.0307\cdot
2^{1/3}]^{-1}$ as in (\ref{e:3a}).  Then $\BS{\gA}$ is non-negative.
\end{theorem}

Numerically, this gives 
$$
Z\leq 56
$$ 
when evaluated with the
experimental value $\alpha \approx 1/137.037$ for the fine structure
constant. Alternatively, considering hydrogen only, i.e., $Z=1$, we
obtain the upper bound 
$$
\alpha \leq 1/8.139
$$ 
for the fine structure constant. It is a challenge to improve
this result so that it covers all physical nuclear charges
and the physical value of the fine structure constant, as was done
for $K=N=1$ and $\gA=\nul$ in \cite{Evansetal1996}.
\medskip

\begin{proof} 
  The first step in our proof is to utilize (\ref{e3}) to replace
  $V_c$ by the one-body operator $(-1/\tilde \alpha_c) \sum_{\nu=1}^N
  |-i\nabla_\nu+e\gA|$, where $\tilde \alpha_c$ is given by
  (\ref{e:3a}) with $q=2$, as we explained just after (\ref{e:3a}).
  (The idea of using the relativistic stability result (\ref{e3}) to
  bound the Coulomb potential by a one-body operator first appears in
  \cite{Liebetal1995}.)  Our energy $\BS{\gA}$ is now bounded below by
\begin{equation}\label{E}
\cE'(\Psi) := 
(\Psi, \sum_{\nu=1}^N (D_\nu(\gA) - \kappa
|-i\nabla_\nu +e\gA(\gr_\nu)|)\Psi) +
\frac1{8\pi}\int_{\rz^3}B(\gr)^2d\gr,
\end{equation}
where $\kappa := \alpha/\tilde\alpha_c$.

The first term on the right side of (\ref{E}) is bounded below by the
sum of the negative eigenvalues, $- \tr h_-$, of the one-body operator
$$
h:= \Lambda^+\Bigl( D(\gA) - \kappa 
|-i\nabla +e\gA(\gr)|\Bigr) \Lambda^+,
$$
where $\Lambda^+$ is the projector onto the positive spectral subspace of
$D(\gA)$.

Let us define
$$S:=|D(A)|-\kappa |-i\nabla + e\gA (\gr)|,$$
whence  $h=\Lambda_+\ S\Lambda_+$, because $\Lambda_+ D(\gA)\Lambda_+
=\Lambda_+ |D(\gA)|\Lambda_+ $. 
We note that for any two
self-adjoint operators $X$ and $\rho$ with $X\ge 0$ and
$0\le\rho\le 1$, $\tr X\ge \tr \rho X$. With $\rho$ taken to be
the projector onto the negative spectral subspace of $h$ we then have
that

\begin{eqnarray}
\tr  h_- &= & -\tr  \rho h =-\tr  \rho \Lambda_+ S\Lambda_+\nonumber\\ 
       &= & \tr  \rho \Lambda_+ S_- \Lambda_+ - \tr  \rho\Lambda_+ S_+ 
       \Lambda_+\label{eq:12}\\ \nonumber
       &\le & \tr  \Lambda_+ S_- \Lambda_+.
\end{eqnarray}

We introduce the $4\times4$ unitary 
$$U=\pmatrix{ 0&1\cr 
             -1&0}$$
and note that $U^{-1} D(\gA) U=-D(\gA)$. Therefore,
$U^{-1}\Lambda_+U=\Lambda_-$.

It follows from the spectral theorem that for any self-adjoint
$X$, unitary $U$, and function $F$
$$F(U^{-1} X U)= U^{-1}F(X) U.$$

With $F(t)=|t|$, we then have that $U^{-1}|D(\gA)|U=|D(\gA)|$,
and hence
$U^{-1} S\,U=S$.

Therefore, since $U^{-1}\Lambda_+U=\Lambda_-$, and with
$F(t)={1\over 2}$ $(|t|-t)=t_-$, we have that $U^{-1}S_-U=S_-$ and 
$$\tr  \Lambda_+S_- \Lambda_+ = \tr  \Lambda_- S_-\Lambda_-.$$

Hence, using (\ref{eq:12}),

$$
 \tr  h_- \leq{1\over 2} \tr  (\Lambda_+ S_- \Lambda_+) + {1\over 2} 
 \tr  \Lambda_-S_-\Lambda_-  
 ={1\over 2} \tr  S_- .$$
(Note: much of the preceding discussion was needed only to get the
factor 1/2 here. This factor improves our final constants for
stability.)

Next, we use the BKS inequality (\ref{e2}) to bound $\tr S_-$ as follows:

\begin{equation}
        \tr h_-\leq {1\over 2} \tr  S_- \le {1\over 2} \tr  \Bigl[D(\gA)^2 - 
        \kappa^2 | - i\nabla +e\gA(\gr)|^2\Bigr]_-^{1/2}.
\end{equation}

However, $D(\gA)^2 = \pmatrix{Y&0\cr 0&Y}$  with $Y=P(\gA)+m^2$, and
where 
$$
P(\gA)=\left[\mbox{\boldmath$\sigma$} \cdot(-i \nabla  +e\gA)\right]^2
=|-i \nabla + e\gA(\gr)|^2 +e\mbox{\boldmath$\sigma$}\cdot \gB(\gr)$$
is the Pauli operator.

Since $X\mapsto\tr  X_-^{1/2}$ is operator monotone decreasing,
we see that our lower bound for the energy is monotone increasing in
$m$, and thus it suffices to prove the positivity of $\bHN{\gA} $ in
the massless case. The key observation is that our lower bound involves
only $\tr S_-$ in the entire one-body space, not the positive energy
subspace. The energy would not be obviously monotone in $m$ if we had
to restrict functions to the positive subspace, since changing $m$
would also entail changing the space.  This problem does not arise in
the absence of the positive subspace constraint.

  Because of the `diagonal' structure of the operator $S$, we can drop
  the factor $1/2$ by replacing the trace on $L^2(\rz^3)\otimes\cz^4$
  by the trace on $L^2(\rz^3)\otimes\cz^2$. This yields
\begin{eqnarray} &&\BS{\gA}[\Psi, \Psi] \geq -\tr[P(\gA) -
  \kappa^2 (-i\nabla+e\gA)^2]_-^{1/2} + \frac1{8\pi}\int_{\rz^3}
  B(\gr)^2d\gr \nonumber \\ &\geq& -2\tr[(1-\kappa^2)(-i\nabla+e\gA)^2-
  e B]_-^{1/2} + \frac1{8\pi}\int_{\rz^3} B(\gr)^2d\gr. \label{e:5}
\end{eqnarray} We regard the operator in the second line as acting on
functions (of one component only) instead of spinors, which accounts
for the factor two (and not one)..  Finally we apply the Lieb-Thirring
inequality (\ref{e4}) to the right hand side yielding (recall that
$e^2=\alpha$) $$ \BS{\gA}[\Psi, \Psi]
  \geq[-2L_{\frac12,3}\, \alpha(1-\alpha^2/\tilde\alpha_c^2)^{-3/2}
  +1/(8\pi)]\int_{\rz^3} B^2(\gr)^2d\gr.  $$
  Thus we need \begin{equation}\label{alphac}
 (16\pi
  L_{\frac12,3}\alpha)^{2/3} \leq 1-\alpha^2/\tilde\alpha_c^2.
\end{equation} Since the right hand side of this inequality is monotone
decreasing in $\alpha$ for positive $\alpha$, while the left hand side
is monotone increasing, there is a unique $\alpha_c$ for which equality
holds in (\ref{alphac}).  Inserting the value (\ref{e:3a}) with $q=2$
for $\tilde\alpha_c$ yields---together with the second requirement on
$Z_1,...,Z_K$ in the relativistic bound---the claimed stability
criterion.  \end{proof}

\section{Instability with the Free Dirac Operator  \label{s3}}

In this section we shall discuss the Brown-Ravenhall model
(\cite{BrownRavenhall1951}). That is to say we consider the energy
expression (\ref{B}) with zero vector potential in the definition of
the allowed electronic states (\ref{HS}), i.e., we take only
$\Psi\in\bHN{\nul}=\atp^N\cH$, where $\cH$ is the positive spectral
subspace of the operator $-\balpha\cdot i\nabla + m \beta$. We shall
prove that there is no stability is in this model by showing that for
any (sufficiently large) particle number, $N$, and any $\alpha >0$ the
energy is unbounded below. In other words, ``stability of the first
kind'' is violated. It is nevertheless true, however, that for any
choice of particle numbers and nuclear charges there is always a
sufficiently small, nonzero $\alpha$ such that the energy {\it is}
bounded below by zero.

Since the positive spectral subspace $\cH$ for the free Dirac operator
is {\it not} invariant under gauge transformations we see that this
Brown-Ravenhall model is not gauge invariant. (The previous, modified
model discussed in Section \ref{s2} is not only stable, it is also
gauge invariant.) More precisely, the energy spectrum depends not only
on $\gB$ but in fact on the full gauge potential $\gA$.  The
Brown-Ravenhall model is therefore physically meaningfully defined only
if we make a fixed choice of gauge. The natural choice is the Coulomb
gauge (radiation gauge), $$ \nabla\cdot\gA=0, $$ since in quantum
electro-dynamics this gauge implies that electrons interact via the
usual Coulomb potentials and the coupling to the transverse  field
is minimal, i.e.,  derivatives are replaced by covariant derivatives.

The interesting quantity is the lowest energy that the system can have.
\begin{definition}[Energy]\label{def:energy}
$$ 
E_{N,K}:= \inf  \BS{\nul}[\Psi,\Psi].
$$ 
where the infimum is taken over all divergence free $\gA$
        fields, pairwise distinct nuclear locations $\gR_1,...,\gR_K$ and
        normalized, antisymmetric states $\Psi \in \bHN{\nul}$.
        
\end{definition}

\subsection{Stability with Small $\alpha$ and Small Particle Number 
  \label{ss31}}

Since this result is not a main point of this paper we shall be
brief---even sketchy.  If a single particle $\Psi $ is in the positive
spectral subspace of $D(\nul)$ then the action of $D(\nul)$ on $\Psi $
is the same as multiplication of each component by $(p^2 + m^2)^{1/2}$
in Fourier space.  For such functions we see that $(\Psi, D(\nul)
\Psi)$ exceeds $(\Psi, |\nabla|\Psi)$, so we may as well replace
$D(\nul)$ by $|\nabla|$ and also drop the condition that $\Psi$ belong
to the positive spectral subspace of $D(\nul)$.

The next step is to use the lower bound on $V_c$ in (\ref{e3}) so that the
energy is now bounded below by a sum of one-body operators, in a
manner similar to that in Section \ref{s2} with $\tilde\alpha_c$ as in
Theorem \ref{t1}):
\begin{equation}\label{e:ss1}
  \BS{\nul}[\Psi,\Psi] \geq (\Psi, \sum^N_{\nu=1} (1-
  \frac{\alpha}{\tilde\alpha_c})|\nabla_\nu| \Psi)+
  \sqrt{\alpha}\int_{\rz^3} \gj \cdot \gA +
  \frac1{8\pi}\int_{\rz^3}B^2.
\end{equation}
(Note again that the `effective spin' $q$ is $2$, as can be seen be
repeating the above argument.) Here $\gj(\gr)$ is the current in the
state $\Psi$ and it is trivially bounded above pointwise by the
density $\rho(\gr)$ in the state $\Psi$ (defined in Appendix
\ref{a:2}). Therefore the integral involving $\gA$ is bounded below by
$$ \int_{\rz^3}\gj\cdot\gA \geq-\int_{\rz^3}\rho A
\geq-\|A\|_6N^{1/3}\|\rho\|_{4/3}^{2/3}.
$$
Now $\int B^2 \geq \int |\nabla A |^2$ and this is not less than
$K_3^{-2} \|A\|_6^2$ by Sobolev's inequality where $K_3 =
4^{1/3}(3\pi)^{-1/2}\pi^{-1/6}$ (see \cite{Lieb1983}, p.~367).
Similarly, the kinetic energy $(\Psi,\sum^N_{\nu=1}|\nabla_\nu|\Psi)$
is bounded below by $1.63q^{-1/3}\int\rho^{4/3}$, which was proved
by Daubechies \cite{Daubechies1983} and which follows from
(\ref{e:4a}).  If we use these inequalities and then minimize the
energy with respect to the unknown quantity $\|A\|_6$, we easily find
that the energy is non-negative as long as
$$ 
1.63 (1-\alpha/\tilde\alpha_c) \geq 2\pi N^{2/3}K_3^2q^{1/3}\alpha
$$ with $q=2$. 

We shall show in Subsection \ref{ss32} that the condition that
$N^{2/3}\alpha$ is small, which---as we just proved---ensures
boundedness from below, is in fact also necessary for the energy to be
bounded from below.

\subsection{ Instability for All $\alpha$ and Large Particle
Number \label{ss32}}

The main result of this section is that there is no stability in this
model for any fixed, positive $\alpha$ if $N$ and $K$ are allowed to
be arbitrary.
\begin{theorem}[Instability]
  There exists a universal number $C>0$ such 
  that for all values of the parameters $\alpha >0$, $m \geq 0$,
  $K=1,2,3,\ldots $, and all values of $N=1,2,3,\ldots$, and of
  $Z_1,Z_2,\ldots,Z_K$ satisfying
  $$\sum_{\kappa=1}^K Z_\kappa> C\max\{\alpha^{-3/2},1\},
  \ \ N >
  C\max\{\alpha^{-3/2},1\},\ 
  \ \ \sum_{\kappa=1}^K Z_\kappa^2>2$$
  we have
$$E_{N,K}=-\infty.$$
\end{theorem}
 
\begin{proof}
The theorem follows if, for all $E>0$, we show the existence of 
three quantities for which $\BS{\nul}[\Psi,\Psi]\leq -E$,
with $\Psi=\psi_1\wedge\cdots\wedge\psi_N$:
\begin{description}
\item{A.} \ A vector potential $\gA$ with $\nabla\cdot\gA=0$.
\item{B.} \ Orthonormal spinors $\psi_1,\ldots,\psi_N\in\cH$.
\item{C.} \ Nuclear coordinates $\gR_1,\ldots,\gR_K$. 
\end{description}

Our construction will depend on four parameters (to be specified at
the very end), $\delta>0$ a momentum scale, which we shall let tend to
infinity, $\theta>0$, which will be chosen sufficiently small (but
independently of $N$), and $P,A_0>0$ which will be chosen as functions
of $N$.  Finally we denote by $\gn_1,\gn_2,\gn_3$ the 
coordinate vectors $(1,0,0),(0,1,0),(0,0,1)$ respectively.
We shall use the notation that $\bomega_p=\gp/|\gp|$ is the unit
vector in the direction $\gp\in\rz^3$.

{\it A. The vector potential.\/} We choose the vector potential $\gA$
to have Fourier transform
$$ \hat{\gA}(\gp) := A_0\chi_{B(0,5\delta)}(\gp)
\left(\gn_2\cdot\bomega_\gp\right)
\gn_3\times
\bomega_\gp,
$$ where $\chi_{B(0,5\delta)}$ denotes the characteristic function (in
$\gp$-space) of the ball $B(0,5\delta)$ centered at $0$ with radius
$5\delta$.  Note first that $\gA$ is real since $\hat{\gA}$ is real
and $\hat{\gA}(\gp)=\hat{\gA}(-\gp)$.
Moreover, $\gA$ is divergence free, i.e., it  is
in the Coulomb gauge, since
$-i\widehat{\nabla\cdot\gA}(\gp)=\gp\cdot\hat{\gA}(\gp)=0$.  We easily
estimate the self-energy of the magnetic field $\gB=\nabla\times\gA$
corresponding to $\gA$
\begin{equation}\label{eq:umagnetic}
  \frac{1}{8\pi}\int (\nabla\times\gA)^2
  =\frac{1}{8\pi}\int_{\rz^3}|\gp\times \hat\gA(\gp)|^2d\gp \leq
  A_0^22^{-1}\int_0^{5\delta} p^4dp =2^{-1}5^4A_0^2\delta^5.
\end{equation}
Finally, we note for later use that
\begin{equation}\label{eq:later}
        \hat{\gA}(\gp)\cdot \gn_1=-A_0\chi_{B(0,5\delta)}(\gp)
\left(\gn_2\cdot\bomega_\gp\right)^2.
\end{equation}

{\it B. The orthonormal spinors.\/}
For $\gp_0\in\rz^3$ define 
\begin{equation}\label{eq:udef}
 u_{\gp_0}(\gp) = \sqrt{3/(4\pi)}\delta^{-3/2}
\left(\begin{array}{c} \chi_{B(0,\delta)}(\gp-\gp_0)\\ 
  0\end{array}\right).
\end{equation}
We then have a normalized $\psi_{\gp_0}\in\cH$ given by 
$$ \widehat{\psi_{\gp_0}}(\gp)=\left(2E(p)(E(p)+E(0))\right)^{-1/2}
\left(\begin{array}{c} (E(p)+E(0))u_{\gp_0}(\gp) \\ 
  \gp\cdot\bsigma u_{\gp_0}(\gp)
        \end{array}\right),
$$ 
where $E(p)=(p^2+m^2)^{1/2}$.  Recall that this is the
general form of a spinor in the positive spectral subspace $\cH$ for
the free Dirac operator.

For the sake of simplicity we shall henceforth assume that $m=0$. We
leave it to the interested reader to check the estimates for the
general case $m\ne0$. We shall indeed consider spinors with momenta
$p$ such that we have $p^2(m^2)^{-1}\to\infty$ as
$\delta\to\infty$, i.e., $E(p)\approx p$. It is therefore
straightforward to estimate the expressions in the general case
$m\ne0$ by the corresponding expressions for $m=0$. 

In particular, we have, for $m=0$,
$$ \widehat{\psi_{\gp_0}}(\gp)=2^{-1/2} \left(\begin{array}{c}
u_{\gp_0}(\gp) \\ \bomega_\gp\cdot\bsigma u_{\gp_0}(\gp)
\end{array}\right),
$$ We shall choose $N$ points $\gp_1,\ldots,\gp_N\in\rz^3$ such that
the following conditions are satisfied.
\begin{enumerate}
\item \label{it:dist} $\min_{\nu\ne \mu}|\gp_\nu-\gp_\mu|>2\delta$
\item \label{it:P} $P \leq p_\nu\leq 2P$, for all $\nu=1,\ldots,N$
\item \label{it:angle} $\bomega_{\gp_\nu}\cdot \gn_1\geq
  1-\theta^2$, for all $\nu=1,\ldots,N$
\end{enumerate}
Condition~\ref{it:dist} ensures that the spinors
$\psi_{\gp_1},\ldots,\psi_{\gp_N}$ are orthonormal. The importance of
Conditions~\ref{it:P} and \ref{it:angle} will hopefully become clear
below.

In order that Conditions~\ref{it:dist},\ref{it:P} and \ref{it:angle}
are consistent with having $N$ points (for large $N$) we 
must ensure that $N$ balls of radius $\delta$
can be packed into the domain defined by Condtions~\ref{it:P} and 
\ref{it:angle}. Since small enough balls can fill at least half the volume 
of the given region we simply choose $P$
such that
$$ 2N\leq\frac{\mbox{Vol}\left(\{\gp\ |\ P\leq p \leq 2P,
\ 1-\theta^2\leq
  \bomega_{\gp}\cdot \gn_1\}\right)}{(4\pi/3)\delta^3}
=\frac{7}{2}\theta^2\frac{P^3}{\delta^3}.
$$ 
Note that the assumption that $N$ is larger than some universal 
number ensures that the balls are small, i.e., that
$\delta$ is small enough compared to $P$. 
Thus, we have the condition
\begin{equation}\label{eq:PNcondition}
  P\geq \left(\frac{4N}{7\theta^2}\right)^{1/3}\delta.
\end{equation}
In particular, since we shall choose $\theta$
independently of $N$ we may assume that $N$ is large enough that
the above condition implies
$P\theta\geq 2\delta$.
(Since we shall choose $\delta\to\infty$ we see that the momenta of
the spinors satisfy $p^2m^{-2}\to\infty$.)
        
We are now prepared to calculate
$(\Psi,\sum_{\nu=1}^ND_\nu(\gA)\Psi)$, where
$\Psi=\psi_{\gp_1}\wedge\cdots\wedge\psi_{\gp_N}$. We obtain
\begin{equation}\label{eq:HNexpectation}
  (\Psi,\sum_{\nu=1}^ND_\nu(\gA)\Psi)= \sum_{\nu=1}^N\left(T_\nu +
  \int e\gj_\nu\cdot\gA\right),
\end{equation}
where 
\begin{equation}\label{current}
\gj_{\nu}(\gr):=\psi_{{\gp_{\nu}}}^*(\gr)\balpha\psi_{{\gp_{\nu}}}(\gr),
\end{equation}
is the current of the $\nu$-th one-electron state
$\psi_{{\gp_{\nu}}}$, and
\begin{equation}\label{eq:Tn}
  T_\nu:=(\psi_{\gp_{\nu}}, (- i\balpha\cdot\nabla+\beta m)
  \psi_{\gp_{\nu}}) =\int E(p)|u_{\gp_{\nu}}(\gp)|^2d\gp\leq 
  p_{\nu}+\delta,
\end{equation}
since we have assumed that $m=0$ and hence $E(p)= p$.

We must evaluate the current integral
\begin{eqnarray}
        \int \gj_{\nu}\cdot\gA
        &=&(2\pi)^{-3/2}2\Re\iint u_{\gp_{\nu}}^*(\gq-\gp)(\hat{\gA}(\gp)
        \cdot\bsigma)
        (\bomega_\gq\cdot\bsigma)u_{\gp_{\nu}}(\gq) d\gp d\gq\nonumber\\
        &=&(2\pi)^{-3/2}
        2\Re\iint\!\!\!\!\!\begin{array}[t]{rl}
        \Bigl[\!\!\!\!\!&\hat{\gA}(\gq-\gp)\cdot\bomega_\gq
        u_{\gp_{\nu}}^*(\gp)
        u_{\gp_{\nu}}(\gq)\\
        &
        +iu_{\gp_{\nu}}^*(\gp)\left(\hat{\gA}(\gq-\gp)
        \times\bomega_\gq
        \right)\cdot\bsigma
        u_{\gp_{\nu}}(\gq) \Bigr]d\gp d\gq.\end{array}
        \label{eq:currentdefinition}
\end{eqnarray}
We first observe that 
\begin{eqnarray*}
        \lefteqn{2\Re\iint\Bigl[iu_{\gp_{\nu}}^*(\gp)\left(\hat{\gA}(\gq-\gp)
        \times\bomega_\gq
        \right)\cdot\bsigma
        u_{\gp_{\nu}}(\gq) \Bigr]d\gp d\gq}&&\\
        &=& 2\Re\iint\Bigl[iu_{\gp_{\nu}}^*(\gp)\left(\hat{\gA}(\gq-\gp)
        \times\bomega_\gq
        \right)\cdot\gn_2\sigma_2
        u_{\gp_{\nu}}(\gq) \Bigr]d\gp d\gq=0.
\end{eqnarray*}
The terms containing $\sigma_1$ and $\sigma_3$ vanish as they are
clearly imaginary. The term with $\sigma_2$ vanishes
beacuse of the choice (\ref{eq:udef}) of $u_{\gp_{\nu}}$.

Note that 
$u_{\gp_{\nu}}^*(\gp)u_{\gp_{\nu}}(\gq)=0$ unless
$|\gp-\gq|<2\delta$ and $|\gq-\gp_\nu|\leq\delta$.
Thus $\bomega_{\gp_\nu}-\bomega_{\gq}=p_\nu^{-1}(\gp_\nu-\gq)+
\bomega_\gq p_\nu^{-1}(q-p_\nu)$ and we obtain for 
$|\gp_\nu-\gq|<2\delta$ that
$$
        \left|\bomega_{\gp_\nu}-\bomega_{\gq}\right|\leq2\delta p_\nu^{-1}\leq
        2\delta P^{-1},
$$
where we used that $p_\nu\geq P$. Since $\bomega_{\gp_\nu}\cdot\gn_1
>1-\theta^2$ we have that $|\bomega_{\gp_\nu}-\gn_1|\leq\theta$ and 
hence
$$
        \left|\hat{\gA}(\gq-\gp)\cdot\left(\bomega_\gq-\gn_1\right)\right|
        \leq \left(2\delta P^{-1}+\theta\right) A_0\leq 2\theta A_0.
$$
Hence, since $|\gp-\gq|<2\delta$ we get from (\ref{eq:later})
that
$$
        \hat{\gA}(\gq-\gp)\cdot\bomega_\gq \leq
        -A_0\left[\left(\bomega_{\gq-\gp}\cdot\gn_2\right)^2
        -2\theta\right].
$$
Thus, 
\begin{eqnarray*}
        \int \gj_{\nu}\cdot\gA&=&(2\pi)^{-3/2}2\Re\iint
        \hat{\gA}(\gq-\gp)\cdot\bomega_\gq
        u_{\gp_{\nu}}^*(\gp)
        u_{\gp_{\nu}}(\gq)d\gq d\gp\\
        &\leq&-
        \frac{3}{(2\pi)^{5/2}}A_0\delta^{-3}\iint_{|\gq|,|\gp|<\delta}
        \left[\left(\bomega_{\gq-\gp}\cdot\gn_2\right)^2
        -2\theta\right]d\gp d\gq\\
        &=&\frac{3}{(2\pi)^{5/2}}A_0\delta^{3}\iint_{|\gq|,|\gp|<1}
        \left[\left(\bomega_{\gq-\gp}\cdot\gn_2\right)^2
        -2\theta\right]d\gp d\gq.
\end{eqnarray*}
We now make the choice
$$
        \theta=\frac{1}{3}\left(\iint_{|\gq|,|\gp|<1}1d\gp d\gq\right)^{-1}
        \iint_{|\gq|,|\gp|<1}
        \left(\bomega_{\gq-\gp}\cdot\gn_2\right)^2d\gp d\gq
$$
and arrive at
$$
        \int \gj_{\nu}\cdot\gA\leq -\frac{4}{3}(2\pi)^{-1/2}A_0\theta\delta^3.
$$

{F}rom (\ref{eq:HNexpectation}) and (\ref{eq:Tn}) we therefore obtain
\begin{equation}\label{eq:HNestimate}
        (\Psi,\sum_{\nu=1}^ND_\nu(\gA)\Psi)\leq
        \sum_{\nu=1}^N \left[|\gp_{\nu}|+\delta-
        \frac{4}{3}(2\pi)^{-1/2}A_0e \delta^{3}\theta\right].
\end{equation}

{\it C. The nuclear coordinates.\/} 
Finally, we show how to choose the
nuclear coordinates following an idea in \cite{LiebLoss1986}.
Consider the electronic density of the state $\Psi$,
$\rho(\gr)=\sum_{\nu=1}^N|\psi_{\gp_{\nu}}(\gr)|^2$ then
$$ (\Psi,V_c\Psi)\leq -\sum_{k=1}^K\int
\frac{Z_\kappa\rho(\gr)}{|\gr-\gR_k|} d\gr+D(\rho,\rho)
+\sum_{\kappa,\lambda=1 \atop \kappa<\lambda}^K
\frac{Z_\kappa Z_\lambda}{|\gR_\kappa-\gR_\lambda|},
$$ were we introduced
$D(\rho,\rho):=\frac12\iint\rho(\gr)|\gr-\gr'|^{-1}\rho(\gr')d\gr
d\gr'$.

Note now that $\int N^{-1}\rho=1$, i.e., 
$N^{-1}\rho$ can be considered a probability distribution.
We may therefore average $(\Psi,V_c\Psi)$ considered as a function 
of $\gR_1,\ldots, \gR_K$ with respect to the probability measure
$$ 
\gR_1,\ldots, \gR_K\mapsto N^{-1}\rho(\gR_1)\cdots N^{-1}\rho(\gR_K).
$$
We obtain
\begin{eqnarray*}
&&\int (\Psi,V_c\Psi)N^{-1}\rho(\gR_1)\cdots N^{-1}\rho(\gR_K)
d\gR_1\cdots d\gR_K\\
&=& [(1-(Z/N))^2- N^{-2}\sum_{\kappa=1}^KZ_\kappa^2] D(\rho,\rho).
\end{eqnarray*} 
We shall prove that $[...]<0$. There are two cases.

1) $N\geq Z$: By moving electrons to infinity we may assume that
$Z\leq N <Z+1$. Therefore
$$ 
[(1-(Z/N))^2- N^{-2}\sum_{\kappa=1}^K Z_\kappa^2]\leq Z^{-2}-
(Z+1)^{-2}\sum_{\kappa=1}^K Z_\kappa^2   \leq Z^{-2}- 2 (Z+1)^{-2}<0.
$$

2) $N <Z$: We may move nuclei to infinity and assume that
$N<Z<N+ \max_\kappa Z_\kappa$. Therefore
$$       [(1-(Z/N))^2- N^{-2}\sum_{\kappa=1}^K Z_\kappa^2]\leq
(\max_\kappa Z_\kappa)^2 N^{-2} -N^{-2}\sum_{\kappa=1}^K Z_\kappa^2\leq 0.
$$

We can therefore find nuclear positions $\gR_1,\ldots,\gR_K$ such that
$(\Psi,V_c\Psi)\leq0$.

Using these coordinates together with (\ref{eq:umagnetic}) and
(\ref{eq:HNestimate}) we get
$$
  \BS{\nul}[\Psi,\Psi]\leq \sum_{\nu=1}^N
  \left[p_{\nu}+\delta- \frac{4}{3}(2\pi)^{-1/2}A_0e
    \delta^{3}\theta\right]+2^{-1}5^4A_0^2\delta^5.
$$
We now choose $P=\left(\frac{4N}{7\theta^2}\right)^{1/3}\delta$ 
in accordance with
(\ref{eq:PNcondition}).  We then choose $A_0$ such that 
$\frac{4}{3}(2\pi)^{-1/2}A_0e
\delta^{3}\theta=4P$, i.e.,
$$ A_0=\frac{3}{112^{1/3}}(2\pi)^{1/2}\theta^{-5/3}e^{-1}
        \delta^{-2}N^{1/3}.
$$ Then, using Condition~\ref{it:P} we have $ p_{\nu}-
(8\pi/3)A_0e \delta^{3}\theta\leq -2 P$.  If we insert these
values and estimates above we find
$$ \BS{\nul}[\Psi,\Psi]\leq\delta\left[-2N^{4/3}
\left(4/(7\theta^2)\right)^{1/3}+N+
3^25^4 112^{-2/3}\pi \theta^{-10/3}\alpha^{-1}
N^{2/3}\right].
$$
It is now clear that 
the expression in $[\ ]$ will be negative for $N$ sufficiently large. 
The energy can therefore be made arbitrarily negative by choosing 
$\delta$ large. 
\end{proof}

\begin{appendix}

\section{BKS Inequalities\label{a:1}}
As a convenience to the reader we give a proof of some cases of the
inequalities due to Birman, Koplienko, and Solomyak
\cite{Birmanetal1975}. The case needed in Section \ref{s2} corresponds
to $p=2$ below. There we are interested in $(B-A)_-$, but here we
treat $(B-A)_+$ to simplify keeping track of signs. The proof is the
same.  Recall that $X_+:= (|X|+X)/2$.
\begin{theorem}\label{t:spur}
  Let $p\geq 1$ and suppose that $A$ and $B$ are two nonnegative,
  self-adjoint linear operators on a separable Hilbert space such that
  $ (B^p - A^p)_+^{1/p}$ is trace class. Then $(B-A)_+$ is also trace
  class and
$$
\tr  (B-A)_+ \leq \tr (B^p-A^p)_+^{1/p}.
$$
\end{theorem}
\begin{proof}
  Our proof will use essentially only two facts: $X\mapsto X^{-1}$ is
  operator monotone decreasing on the set of nonnegative self-adjoint
  operators (i.e., $X\geq Y \geq 0 \Longrightarrow Y^{-1} \geq
  X^{-1}$) and $X\mapsto X^r$ is operator monotone increasing on the
  set of nonnegative self-adjoint operators for all $0<r\leq 1$.
  Consequently, $X\mapsto X^{-r}$ is operator monotone decreasing for
  $0<r\leq 1$.

  As a preliminary remark, we can suppose that $B\geq A$. To see this,
  write $B^p = A^p + D$. If we replace $B$ by $[A^p + D_+ ]^{1/p}$
  then $(B^p-A^p)_+ =D_+$ is unchanged, while $X:=B-A \mapsto [A^p + D_+
  ]^{1/p} -A$ can only get bigger because $X \mapsto X^{1/p}$ is
  operator monotone on the set of positive operators.  Since the trace
  is also operator monotone, we can therefore suppose that $D= D_+$,
  i.e., $B^p = A^p + C^p$ with $A,B, C \geq 0$.  Our goal is to prove
  that
\begin{equation}
\tr \left[ (A^p + C^p)^{1/p} -A \right] \leq \tr \, C, \label{genug}
\end{equation}
under the assumption that $C$ is trace class.

To prove (\ref{genug}) we consider the operator $X := [A^p +
C^p]^{1/p}-A$, which is well defined on the domain of $A$. We assume,
at first, that $A^p \geq \varepsilon^p$ for some positive number
$\varepsilon$. Then, by the functional calculus, and with
$$
E:=[A^p +C^p]^{(1-p)/p}  \quad\quad {\rm and} \quad\quad P:= A^{1-p}- E
$$
we have
\begin{equation}
X= E[A^p + C^p] - A^{1-p}A^p = -PA^p +EC^p  \label{one}
\end{equation}
Clearly, $P\geq 0$ and $0\leq P \leq \varepsilon^{1-p}$.

Let $Y:=EC^p$. We claim that $Y$ is trace class. This follows from
$Y^*Y=C^p E^2 C^p \leq C^p C^{2-2p} C^p =C^2$. Thus, $|Y|\leq C$, and
hence $\tr Y =\tr C^{p/2}EC^{p/2} \leq \tr C$.

It is also true that $P$ is trace class. To see this, use the integral
representation, with suitable $c>0$, $A^{1-p} =c\int_0^{\infty}
(t+A)^{-1} t^{1-p}\, dt$. Use this twice and then use the resolvent
formula. In this way we find that
$$
P=c\int_0^\infty(A^p+t)^{-1}C^p(A^p+C^p+t)^{-1}t^{(1-p)/p}\,dt.
$$ Since $C$ is trace class, so is $C^p$, and the integral converges
because of our assumed lower bound on $A$. Thus, $P$ is trace class
and hence there is a complete, orthonormal family of vectors $v_1,
v_2,\dots $, each of which is an eigenvector of $P$.

Since $X \geq 0$, the trace of $X$ is well defined by
$\sum_{j=1}^{\infty} (v_j, X v_j)$ for {\it any} complete, orthonormal
family. The same remark applies to $EC^p$ since it is trace class.
Thus, to complete the proof of (\ref{genug}) it suffices to prove that
$(v_j, PA^p \ v_j)\geq 0$ for each $j$. But this number is $ \lambda_j
(v_j, A^p \ v_j)\geq 0$, where $ \lambda_j$ is the (nonnegative)
eigenvalue of $P$, and the positivity follows from the positivity of
$A$.

We now turn to the case of general $A\geq0$.  We can apply the above
proof to the operator $A+\varepsilon$ for some positive number
$\epsilon$.  Thus we have
\begin{equation}
  \tr \left[[(A+\varepsilon)^p + C^p]^{1/p} - (A+\varepsilon)\right]
  \leq \tr \, C.  \label{(3)}
\end{equation}
Let $\varphi_1,\varphi_2,\ldots$ be an orthonormal basis chosen from
the domain of $A^p$. This basis then also belongs to the domain of $A$
and the domain of $[(A+\varepsilon)^p + C^p]^{1/p}$ for all
$\varepsilon\geq0$.  We then have
$$
\tr \, X =\sum_j(\varphi_j, X  \varphi_j).
$$ Note that a-priori we do not know that the trace is finite, but
since the operator is non-negative this definition of the trace is
meaningful.  Operator monotonicity of $X^{1/p}$ gives
$$ (\varphi_j,\left[[(A+\varepsilon)^p + C^p]^{1/p} - (A+\varepsilon
)\right] \varphi_j)\geq (\varphi_j, (X - \varepsilon ) \varphi_j).
$$ It therefore follows from (\ref{(3)}), followed by Fatou's Lemma
applied to sums that
\begin{eqnarray*}
  \tr \, C &\geq& \liminf_{\varepsilon\to0}
  \sum_j(\varphi_j,\left[[(A+\varepsilon)^p + C^p]^{1/p} -
  (A+\varepsilon)\right] \varphi_j)\\ 
  &\geq&\sum_j\liminf_{\varepsilon\to0}
  (\varphi_j,\left[[(A+\varepsilon)^p + C^p]^{1/p} -
  (A+\varepsilon)\right] \varphi_j)\\ &\geq& \tr \, X.
\end{eqnarray*}
\end{proof}

\section{Counting Spin States \label{a:2}}

Our goal here is to prove that when $\Psi $ is in $\bHN{\cA} $, the
antisymmetric tensor product of the positive energy subspace of the
Dirac operator (with or without a magnetic field, $\cA$) then the
one-body density matrix is bounded by $2$ and not merely by $4$,
as would be the case if there were no restriction to the positive
energy subspace. This result will allow us to use 2 instead of 4 in
inequalities (\ref{e:3a}) and (\ref{e:5}).  We thank Michael Loss for
the idea of this proof.

The one-body density matrix is defined in terms of an $N$-body
density matrix (or function) by the partial trace over $N-1$ variables.
We illustrate this for functions, but the proof works generally. If 
$\Psi$ is a function, then
$$
\Gamma(\gr,\sigma; \gr', \sigma'):= N \int_{(N-1)}\Psi(\gr, \sigma,
z_2, z_3, \dots, z_N) \overline{\Psi(\gr',\sigma', z_2,\dots , z_N)}
dz_2\cdots dz_N,
$$ 
where $z$ denotes a pair $\gr, \sigma$ and $dz$ denotes integration
over $\rz^3$ and summation over the $q$ `spin' states of $\sigma$. We
are interested in $q=4$, but that is immaterial for the definition.

The kernel $\Gamma$ is trace class; in fact its trace is $qN$. It is
also obviously positive definite as an operator. The first remark is
that $\Gamma \leq 1$ as an operator. To prove this easily, let $\psi$
be any normalized function of one space-spin variable $z$ and define
the function of $N+1 $ variables $\Phi (z_0,\dots, z_N) := \psi
(z_0)\Psi(z_1,\dots, z_N)+ \sum_{j=1}^N (-1)^j \psi(z_j)\Psi (
z_0,\dots, \hat z_j, \dots, z_N)$, where $\hat z_j$ denotes the
absence of $z_j$. This function $\Phi$ is clearly antisymmetric and
the integral over all variables of its square is surely nonnegative.
However, this integral is easily computed (using the normalization of
$\psi $ and $\Psi$) to be $(N+1) - (N+1) (\psi, \, \gamma \psi)$.

The next step is to consider the reduced kernel (without spin) defined
by
$$
\gamma(\gr, \gr'):= \sum_{\sigma =1}^q \Gamma(\gr,\sigma ; \gr', \sigma),
$$ which evidently satisfies the operator inequality $0 \leq \gamma
\leq q$, since $\Gamma \leq 1$.

The electron density referred to in Section \ref{ss31} is defined by
$$
\rho(\gr) := \gamma(\gr, \gr),
$$ but it will not be needed in this Appendix. Another quantity of
interest is the current, defined by
$$ \gj(\gr):= \sum_{\sigma, \tau} \Gamma(\gr,\sigma ; \gr, \tau)
\balpha_{\sigma, \tau}.
$$ 
It follows from this that $|\gj(\gr)| \leq \rho(\gr)$ for every
$\gr\in\rz^3$.

Our goal here is to prove the following fact about $\gamma$:\\ 
{\it If the $N$-body $\Psi $ is in $\bHN{\cA}$ then the corresponding
  $\gamma$ satisfies $0\leq \gamma \leq 2$ as an operator. }

To prove this we introduce the unitary matrix in spin-space (related
 to the charge conjugation operator) 
$$
U= \left(\begin{array}{cc}
 0&1\\ -1&0
\end{array}\right)
$$ 
where $1$ denotes the unit $2\times2$ unit matrix.  With a slight
abuse of notation, we shall also use $U$ to denote the $U\otimes1$
acting on the full one-particle space, i.e., $(U)(\gr,\sigma') =
\sum_\sigma U(\sigma', \sigma)f(\gr, \sigma)$.
The important point to note, and
which is easily verified from the Dirac equation, is that $\psi \in
\cH$ if and only if $U\psi \in \cHm$, the negative spectral subspace of
$D(\cA)$.

Given $f\in L^2(\rz^3)$, we define $F^{\tau}$ to be the spinor $F^\tau
(\gr, \sigma):= f(\gr)\delta_{\sigma,\tau}$.  Then evidently
$(f,\,\gamma f)= \sum_\tau (F^\tau,\, \Gamma F^\tau)$.  However, since
the matrix $U$ merely permutes the spin indices and possibly changes
the sign from $+$ to $-$, we have that $\sum_\tau (F^\tau,\, \Gamma
F^\tau)= \sum_\tau (F^\tau,\, \Gamma_U F^\tau)$, with $\Gamma_U :=
U^{-1}\Gamma U$. (Actually, the proof only requires that $U$ be
unitary, nothing more.)

We claim that $\Gamma + \Gamma_U \leq 1$ in which case we have proved
that $(f,\,\gamma f) \leq q/2 =2$, as claimed.  To see this, we note that
$\Gamma \leq 1$ on $\cH$ and $\Gamma_U \leq 1$ on $\cHm$.
Since the two  subspaces are orthogonal, $\Gamma + \Gamma_U \leq 1$ on
the whole spinor space. 
\end{appendix}

{\sc Acknowledgment: } {\small The authors thank Michael Loss for
  valuable discussions, especially with regard to Appendix \ref{a:2}.
  After we had proved the results in this paper, including the
  inequalities in Appendix \ref{a:1}, Huzihiro Araki kindly informed
  us of the paper by Birman, Koplienko, and Solomyak
  \cite{Birmanetal1975} in which the inequalities of Appendix
  \ref{a:1} were proved 21 years earlier; we are grateful to him for
  this help. We are also greatful to M.~Griesemer for pointing 
  out several errors in the preliminary version of this paper.
  --- The authors also thank the following organizations
  for their support: the Danish Science Foundation, the European
  Union, TMR grant FMRX-CT 96-0001, the U.S. National Science
  Foundation, grant PHY95-13072, and NATO, grant CRG96011.}

%\bibliographystyle{abbrv}
%\bibliography{coulomb}

\end{document}